\documentclass{PoS}

\usepackage{graphicx}
\usepackage{amsmath}
\usepackage{amssymb}
\usepackage{slashed}

\newcommand{\tr}{\rm tr \,}

\newcommand{\CD}{\hat \partial}

\title{On the convergence of chiral expansions \\
for charmed meson masses \\
in the up, down and strange quark masses}

\ShortTitle{On chiral extrapolations for charmed meson masses}

\author{\speaker{Matthias F.\,M. Lutz}\\
        GSI Helmholtzzentrum f\"ur Schwerionenforschung GmbH, \\Planckstra\ss e 1, 64291 Darmstadt, Germany\\
        Technische Universit\"at Darmstadt, D-64289 Darmstadt, Germany\\
        E-mail: \email{m.lutz@gsi.de}}  
\author{{Xiao-Yu Guo}\\
        GSI Helmholtzzentrum f\"ur Schwerionenforschung GmbH, \\Planckstra\ss e 1, 64291 Darmstadt, Germany\\
        E-mail: \email{x.guo@gsi.de}}
\author{Yonggoo Heo\\
        Suranaree University of Technology, Nakhon Ratchasima, 30000, Thailand\\
        E-mail: \email{y.heo@g.sut.ac.th}}       

\abstract{
We discuss the convergence properties of chiral expansions for the pseudoscalar and vector charmed meson masses based on the chiral SU(3) Lagrangian. Conventional expansion strategies as formulated in terms of bare meson masses are shown to suffer from poor convergence properties. This changes once the expansion is set up in terms of on-shell masses. We find a rapid convergence of the chiral expansion from vanishing quark masses up to physical values of the strange quark mass in this case. Detailed results are presented at the one-loop level for the $D$-meson and $D^*$-meson masses. It is emphasized that our results do not depend on the renormalization scale. An approximation hierarchy for the chiral Ward identities of QCD is obtained that keeps the proper form of low-energy branch points and cuts as they are implied by the use of on-shell masses. Given such a scheme we analyzed the charmed meson masses as available on various QCD lattice ensembles. In terms of the determined low-energy constants we consider the coupled-channel interactions of the Goldstone bosons with open-charm mesons. For the isospin violating hadronic decay  width of the $D_{s0}^*(2317)$ we predict the range $(104-116)$\,keV.
}

\FullConference{The 9th International Workshop on Chiral Dynamics\\
		17-21 Sep, 2018\\
		Durham, NC, USA.}

\begin{document}

\section{Introduction}

There is a significant effort to compute charm meson masses on lattice ensembles \cite{Mohler:2011ke,Na:2012iu,Liu:2012zya,Lang:2014yfa,Kalinowski:2015bwa,Moir:2016srx}.
Can this data set help to arrive at stringent predictions of QCD for the coupled-channel  dynamics of open-charm meson systems as accessed in the laboratory? While the computation of 
ground state masses on lattice ensembles is quite matured by now, this is not so much the case for scattering observables as needed for a profound interpretation of experimental data.  
Indeed chiral dynamics predicts the low-energy constants that determine the 
quark-mass dependence of the charmed meson masses to dominate the s-wave coupled-channel interaction of the Goldstone bosons with those charmed mesons \cite{Lutz:2015ejy}. 

While such a link does exist beyond any doubt, it is controversial to what extent it can be 
used in an efficient and reliable manner. The challenges are strange degrees of freedom which are known from phenomenology to very often drive the generation of hadron resonances via coupled-channel dynamics. Thus, any chiral extrapolation in the up and down quark masses only, 
will not be able to do the job. Such a program can be useful only if the role of the strange quark in the chiral Lagrangian can be further clarified. If setup in a conventional manner
a chiral expansion in the strange quark mass appears futile: the convergence properties are quite unfortunate at its physical value.  
In a recent work, this problem has been studied at hand of SU(3) chiral correction terms in the  light baryon masses \cite{Lutz:2018cqo}. It was demonstrated that using on-shell hadron masses in loop contributions does lead to much improved convergence properties of the chiral expansion. 
A power-counting scheme in terms of on-shell hadron masses has been established leading to a convincing convergence pattern. As an unavoidable consequence of such a chiral extrapolation of a hadron mass  non-linear and coupled sets of equations have to be solved. 

In this contribution, we focus on chiral SU(3) expansions of pseudoscalar and vector charmed meson masses. The various facets of the chiral extrapolation challenge are illustrated by an analysis 
of the one-loop expressions, as they are implied by the chiral Lagrangian formulated for the 
charmed meson fields with $J^P = 0^-$ and $J^P= 1^-$ quantum numbers. We report on a successful application of our chiral extrapolation scheme with on-shell hadron masses to the available 
QCD lattice simulation results for the charmed meson masses \cite{Guo:2018kno}.  
Further constraints from lattice results on the s-wave scatterings of $D$-mesons off Goldstone bosons \cite{Liu:2012zya,Moir:2016srx} are considered.

\section{The chiral Lagrangian for open-charm mesons}
\label{sec:2}

The  chiral SU(3) Lagrangian for the ground-state charmed mesons has been constructed in \cite{Kolomeitsev:2003ac,Hofmann:2003je,Lutz:2007sk,Guo:2018kno}, with the anti-triplet fields, $D$ and $D_{\mu \nu}$, of charmed mesons with $J^P= 0^-$ and $J^P = 1^-$ quantum numbers. The terms relevant for the pseudoscalar $D$-meson masses are \cite{Guo:2018kno}
\begin{eqnarray}
&& \mathcal{L}_{}=(\CD_\mu D)(\CD^\mu \bar D)-  M^2\, D \, \bar D
  + 2\,g_P\,\big\{D_{\mu \nu}\,U^\mu\,(\CD^\nu \bar D)
 - (\CD^\nu D )\,U^\mu\,\bar D_{\mu \nu} \big\}
\nonumber\\ 
&& \quad - \,\big( 4\,c_0-2\,c_1\big)\, D \,\bar{D}  \,{\tr} \chi_+ -2\,c_1\,D \,\chi _+\,\bar{D}
 + \, 4\,\big(2\,c_2+c_3\big)\,D\bar{D}\,{\tr} \big(U_{\mu }\,U^{\mu \dagger }\big)- 4\,c_3\, D \,U_{\mu }\,U^{\mu \dagger }\,\bar{D}
\nonumber\\
&& \quad +\,\big( 4\,c_4+2\,c_5\big)\, ({\CD_\mu } D) ({\CD_\nu }\bar{D}) \,{\tr} \big[ U^{\mu }, \,U^{\nu \dagger }\big]_+ /M^2
-2\,c_5\,({\CD_\mu } D) \big[ U^{\mu }, \,U^{\nu \dagger }\big]_+({\CD_\nu }\bar{D}) /M^2\,,
\label{def-kin}
\end{eqnarray}
where
\begin{eqnarray}
&& U_\mu = {\textstyle \frac{1}{2}}\,e^{-i\,\frac{\Phi}{2\,f}} \left(
    \partial_\mu \,e^{i\,\frac{\Phi}{f}} \right) e^{-i\,\frac{\Phi}{2\,f}} \,, \qquad \qquad 
    \Gamma_\mu ={\textstyle \frac{1}{2}}\,e^{-i\,\frac{\Phi}{2\,f}} \,\partial_\mu  \,e^{+i\,\frac{\Phi}{2\,f}}
+{\textstyle \frac{1}{2}}\, e^{+i\,\frac{\Phi}{2\,f}} \,\partial_\mu \,e^{-i\,\frac{\Phi}{2\,f}}\,,
\nonumber\\
&& \chi_\pm = {\textstyle \frac{1}{2}} \left(
e^{+i\,\frac{\Phi}{2\,f}} \,\chi_0 \,e^{+i\,\frac{\Phi}{2\,f}}
\pm e^{-i\,\frac{\Phi}{2\,f}} \,\chi_0 \,e^{-i\,\frac{\Phi}{2\,f}}
\right) \,, \qquad \chi_0 =2\,B_0\, {\rm diag} (m,m,m_s) \,,
\nonumber\\
&& \CD_\mu \bar D = \partial_\mu \, \bar D + \Gamma_\mu\,\bar D \,, \qquad \qquad \qquad \quad \;\; 
\CD_\mu D = \partial_\mu \,D  - D\,\Gamma_\mu \,.
\label{def-chi}
\end{eqnarray}
The quark masses enter via the  $\chi_0$ field (with $m= (m_u+ m_d)/2$) and the octet of the Goldstone bosons is encoded into the $3\times3$ matrix $\Phi$.
The parameter $M$ measures the mass of the $D$ mesons in the chiral limit, provided that a suitable renormalization scheme is applied \cite{Lutz:2018cqo,Guo:2018kno}. 
The 3-point vertex proportional to $g_P$ induces bubble-loop corrections to $D$-meson masses, where 
the hadronic decay width of the $D^*$-meson implies $|g_P| = 0.57 \pm 0.07 $ \cite{Lutz:2007sk}.
The counter terms proportional to $c_0$ and $c_1$ contribute to the $D$-meson masses at both tree and one-loop level. The other counter terms $c_{2-5}$ define their tadpole corrections. The symmetry breaking counter terms involving two $\chi_+$ fields  are not shown in (\ref{def-kin}) but are systematically considered in \cite{Guo:2018kno}. Further terms with the vector $D^*$ fields are listed in \cite{Guo:2018kno}.

The low-energy constant (LEC) do not only contribute to $D$-meson masses but also to subleading order corrections in the scattering processes between the $D$ mesons and Goldstone bosons. The covariant derivative $\CD_\mu$ in the kinetic term of the Lagrangian \eqref{def-kin} generates the leading order two-body chiral interaction, recognized as the Weinberg-Tomozawa term. While this interaction does not modify the $D$-meson masses it provides the leading order contribution to such s-wave scattering processes. Its interaction strength is determined by a single parameter $f$, the chiral SU(3) limit value of the pion-decay constant. At chiral order $Q^2$ the LEC $c_{0-5}$ turn relevant for the scattering processes. Further terms at chiral order $Q^3$ were introduced in \cite{Du:2017ttu},
\begin{eqnarray}
&& \mathcal{L}^{(3)}=
 \, 4\,g_1\,{D}\,[\chi_-,\,{U}_\nu]_- \CD^\nu \,\bar{D}/M
  - 4\,g_2\,{D}\,\big([{U}_\mu,\,[\CD_\nu,\,{U}^\mu]_-]_- + [{U}_\mu,\,[\CD^\mu,\,{U}_\nu]_-]_-\big)\,\CD^\nu\bar{D}/M
\nonumber\\ 
&& \qquad  - \,4\,g_3\,{D}\,[{U}_\mu,\,[\CD_\nu,\,{U}_\rho]_-]_-\,[\CD^\mu,\,[\CD^\nu,\,\CD^\rho]_+]_+\bar{D}/M^3
  + {\rm h.c.}\,.
\label{def-kin}
\end{eqnarray}
While the  LEC $g_i$ in (\ref{def-kin}), do not contribute to the charmed-meson masses, they imply specific contributions to the two-body coupled-channel interaction kernel.

\begin{figure}[t]
\center{
\includegraphics[keepaspectratio,width=0.4\textwidth]{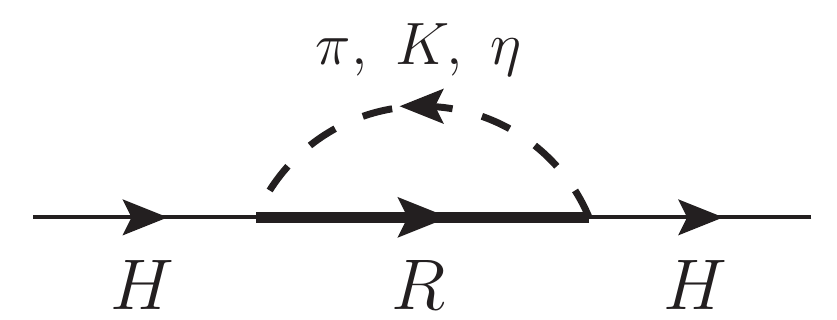} 
}
\vskip-0.2cm
\caption{\label{fig:1} 
The bubble-loop correction to the $D$-meson masses}
\end{figure}

\section{Chiral corrections to the charmed meson masses} 

Given the chiral Lagrangian, the masses of the charmed mesons of type $H$ in either $J^P=0^-$ or $J^P =1^-$ are  determined by the set of coupled and non-linear equations
\begin{align}
	M_H^2 - \Pi_H^{(0)} - \Pi_H^{2-\chi} - \Pi_H^{4-\chi} - \Pi_H^{\rm tadpole} - \Pi_H^{\rm bubble}/Z_H = 0 \,,
	\label{def-Pi}
\end{align}
where the chiral expansion is truncated at the one-loop level.
The term $\Pi_H^{(0)}$ introduces the chiral SU(3) limit of the $D$-meson and $D^*$-meson masses. It is either $M^2$ or $(M + \Delta)^2$ in our notation. The $\Pi_H^{2-\chi}$ and $\Pi_H^{4-\chi}$ contributions are from Lagrangian terms with either one or two $\chi_+$ fields. While the tadpole $\Pi_H^{\rm tadpole}$ is implied by $c_{0-5}$, the bubble $\Pi_H^{\rm bubble}$ is proportional to $g_P^2$. 
The wave-function renormalization factor  $Z_H$ in  (\ref{def-Pi}) is introduced with 
\begin{eqnarray}
	&& Z_H - 1 = \frac{\partial}{\partial M_H^2} \,\Pi_H^{\rm bubble}\, .
	\label{def-Z}
\end{eqnarray}
As was emphasized in  \cite{Guo:2018kno} only with (\ref{def-Z}) it is justified to use a tree-level estimate for $g_P$ in (\ref{def-Pi}).

The bubble function $\Pi_H^{\rm bubble}$ depends on the internal meson masses $m_Q$ (Goldstone boson masses) and $M_R$ ($D$-meson or $D^*$-meson masses) as well as the external charmed-meson mass $M_H$. Note that in a conventional $\chi$PT approach such masses would be replaced by their leading chiral moments, the number of which depending on the target accuracy of the computation. Within dimensional regularization it is straightforward to find expressions for $\Pi_H^{\rm bubble}$. It is convenient to organize such a computation in terms of the Passarino-Veltman reduction scheme \cite{Passarino:1978jh}, where in this case the result is presented in terms of scalar tadpoles, $I_Q$ and $I_R$, and a scalar bubble loop function $I_{QR}$. Such a result is at odds with the expectation from dimensional counting rules. There are various methods how to 
set up renormalization as to have the counting rules realized in a manifest manner. As demonstrated in \cite{Semke:2005sn}, given the Passarino-Veltman scheme it suffices to devise a suitable subtraction scheme for the scalar loop integrals. 

In a realization of this scheme  \cite{Lutz:2018cqo,Guo:2018kno} all terms proportional to a heavy tadpole $I_R$ must be dropped. In addition the renormalized scalar bubble takes the form 
\begin{eqnarray}
&& \bar I_{Q R}  =  -\frac{1 - \gamma^H_{\,R}}{16\,\pi^2} + \frac{1}{16\,\pi^2}\,\frac{m_Q^2}{M_R^2-m_Q^2}\log \frac{m_Q^2}{M_R^2} + \left.\int_{(m_Q+M_R)^2}^\infty \frac{d s}{8\,\pi^2}
\,\frac{p^2}{s^{3/2}}\,\frac{p_{Q R}(s)}{s-p^2} \,\right|_{p^2 = M_H^2}\,,
\nonumber\\
&& p_{Q R}^2(s) = \frac{s}{4}-\frac{M_R^2+m_Q^2}{2}+\frac{(M_R^2-m_Q^2)^2}{4\,s} \,,
\qquad \gamma^H_{\,R} = -  \lim_{m, m_s\to 0}\,\frac{M_R^2-M_H^2}{M_H^2}\,\log \left|\frac{M_R^2-M_H^2}{M_R^2}\right| \,,
\qquad
\label{disp-integral}
\end{eqnarray} 
where we wish to direct the reader's attention to the subtraction term $\gamma_R^H$ in (\ref{disp-integral}). In the limit of an infinite charm quark mass it follows $\gamma_R^H\to 0$. Such a term is required as to arrive at consistent results in the chiral domain with $m_Q < \Delta$ \cite{Lutz:2018cqo}. It is emphasized that the renormalized scalar bubble does not depend on the renormalization scale $\mu$, which enters the result exclusively via the tadpole terms 
\begin{eqnarray}
\bar I_Q = \frac{m_Q^2}{(4\pi)^2} \,\log \frac{m_Q^2}{\mu^2} \,.
\end{eqnarray}
In our scheme  any such contribution proportional to $\bar I_Q$ is absorbed into the tadpole term $\Pi_H^{\rm tadpole}$ with
\begin{eqnarray}
	&& c_2^r =c_2 + \frac{1}{8} g_P^2 \, , \qquad \qquad \qquad
	c_3^r =c_3 - \frac{1}{4} g_P^2 \,.
\end{eqnarray}
As a consequence $\Pi_H^{\rm bubble}$ and $Z_H$ do not depend on the renormalization scale $\mu$. Altogether our mass equation (\ref{def-Pi}) is invariant under any change of $\mu$. This is so since the terms  $\Pi_H^{4-\chi}$ are cast into the unique form such that the $\mu$ dependence from $\Pi_H^{\rm tadpole}$ is balanced exactly \cite{Guo:2018kno}. The rewrite involves the quark masses but also the on-shell meson masses $m_\pi, m_K$ and $m_\eta$.

\begin{figure}[t]
\center{
\includegraphics[keepaspectratio,width=0.49\textwidth]{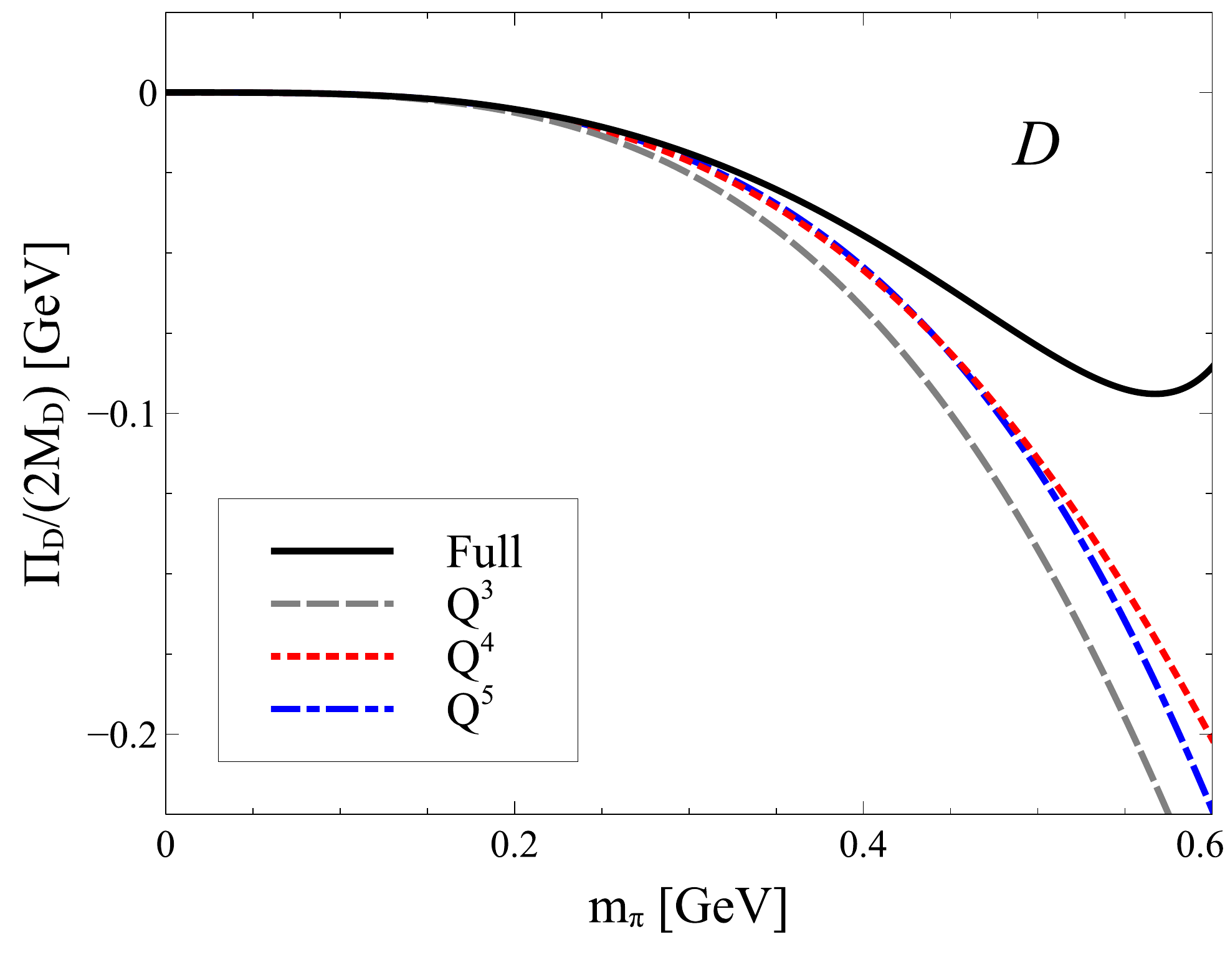}\includegraphics[keepaspectratio,width=0.49\textwidth]{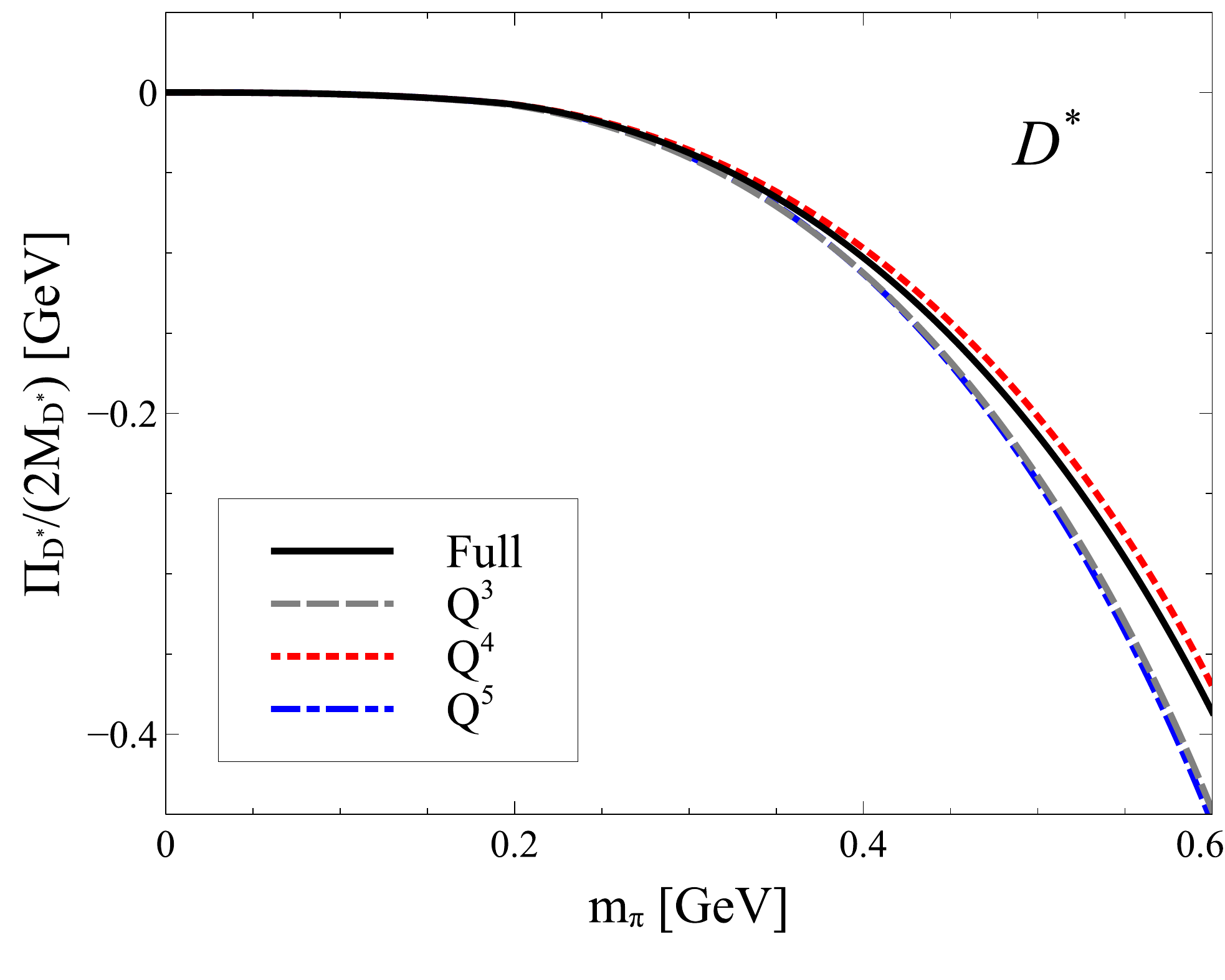} }
\vskip-0.1cm
\caption{\label{fig:2} Charmed meson masses in the flavour limit as a function of the pion mass \cite{Guo:2018kno}. The counting rules (\ref{def-counting-C}) are used.  }
\end{figure}

We now scrutinize chiral expansion strategies of the renormalized bubble loop function. Following
the conventional scheme introduced by Banerjee and collaborators \cite{Banerjee:1994bk,Banerjee:1995wz} the counting rules 
\begin{eqnarray}
 \Delta \sim m_Q \sim Q  \,, \qquad \qquad \Delta_Q = \sqrt{\Delta^2 - m_Q^2} \sim Q\,,\qquad \qquad \frac{\Delta}{M} \sim Q \,,
 \label{def-counting-C}
\end{eqnarray}
are set. Detailed expressions based on (\ref{def-counting-C}) are collected in \cite{Guo:2018kno}. 
In the chiral domain with $m_Q< \Delta$ a further expansion may be applied. It is clear, however, that any expansion that rests on $m_Q < \Delta$ cannot be applicable at the physical point. Despite the attempt to establish a scheme that is applicable at $m_Q \simeq \Delta$, the Fig. \ref{fig:2} illustrates that the counting ansatz (\ref{def-counting-C}) is futile, at least in any application that rests on a few leading order terms.

\begin{figure}[t]
\center{
\includegraphics[keepaspectratio,width=0.95\textwidth]{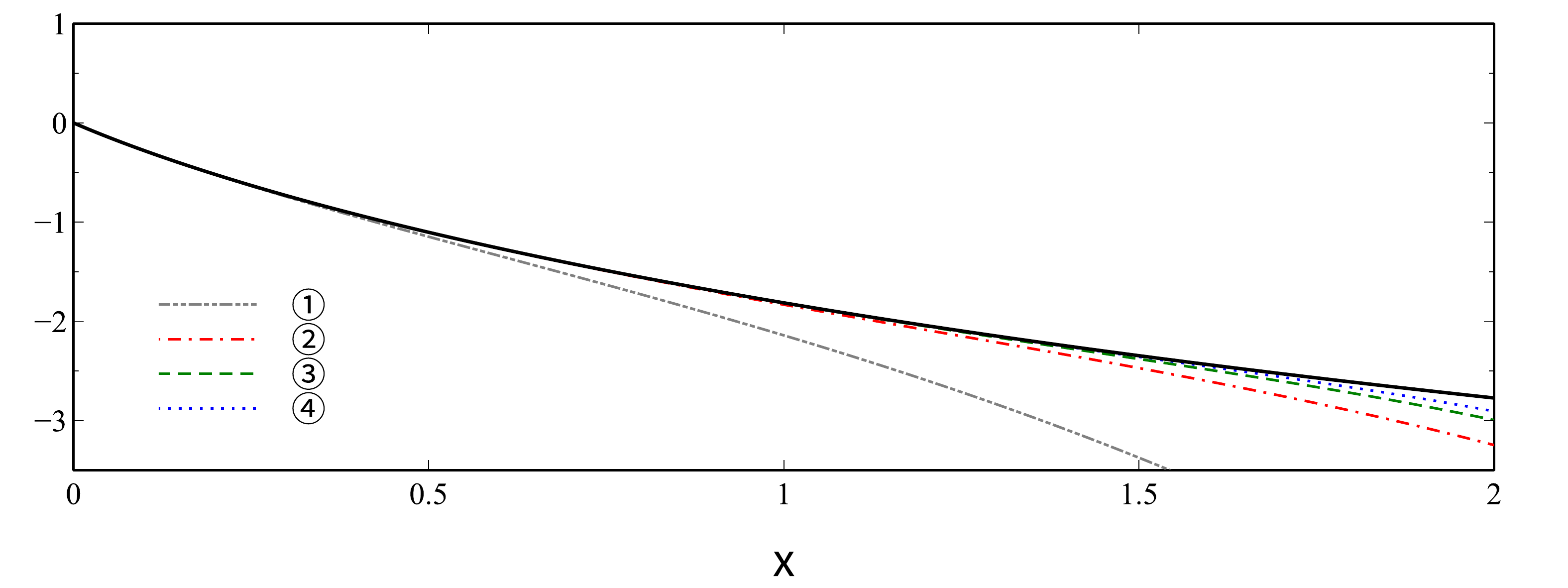}
}
\vskip-0.2cm
\caption{\label{fig:3} The bubble $(4\pi)^2 \bar I_{QR}$, plotted as a function of $x = m_Q/M_H$ with $M_R= M_H$ (solid line). Different broken lines  "1-4" represent different truncations made in the expansions of $f_n$, recorded in \eqref{fn-exp}. }
\end{figure}

How can we overcome this chiral wall? 
Any chiral expansion strategy must deal with a decomposition of the scalar bubble $\bar I_{Q R} $ into its chiral moments. We do so first in the particular case with $M_H = M_R$. In Fig. \ref{fig:3} 
the bubble is plotted as a function of $x=m_Q/M_H$. The analytic structure of this function was 
scrutinized in \cite{Lutz:2018cqo}. 
\begin{align}
	&(4\pi)^2 \bar I_{QR} =-\pi\,\sqrt{x^2}\, f_1(x^2) + x^2\, f_2(x^2) -\frac{1}{2} \,x^2 \,f_3(x^2)\, \log x^2 \,,
	\label{def-IQR}
\end{align}
where each of the functions $f_n(x^2)$ with $f_n(0)=1 $ was shown to be analytic in the circle with $|x| < 2$. 
Thus they can be expanded around $x=0$ within the convergence domain $|x|<2$. The first few moments read

\begin{eqnarray}
&& (4\pi)^2\,\bar I_{QR} = -\Big\{ 1 - \frac{1}{8}\,x^2 - \frac{1}{128}\, x^4 - \frac{1}{1024} \,x^6
+ {\mathcal O} (x^8)\Big\} \,\pi \,\sqrt{x^2}
\nonumber\\
&& \qquad \qquad \;\;\;\;\, +\, \Big\{ 1 - \frac{1}{12}\,x^2 -\frac{1}{120}\, x^4  - \frac{1}{840}\,x^6 + {\mathcal O} (x^8) \Big\} \,x^2
\nonumber\\
&& \qquad \qquad  \;\;\;\;\,
-  \,\frac{1}{2}\,x^2\,\log x^2    \,.
\label{fn-exp}
\end{eqnarray}
A significant cancellation amongst the three terms $f_n$ is observed. This is a consequence of the analytic structure of the bubble loop. Once the correlation in (\ref{fn-exp}) is kept, the expansion converges rapidly up to the Goldstone-boson mass as large as $m_Q = 2\,M_H$. This is illustrated in Fig.~\ref{fig:3}, where the solid line (full result) is confronted with the four cases where each of the $f_n(x^3)$ is approximated by one, two, three or four terms. 

In the standard $\chi$PT approach the correlation (\ref{fn-exp}) is not kept, and in turn the expansion has a sizeable oscillatory part. This is the source of the rather controversial convergence property of a conventional chiral expansion of  hadron masses. A significant improvement is obtained once the expansion is set up in terms of on-shell masses. In this case it is possible to keep such correlations. 
We propose an expansion along the power-counting rules
\begin{eqnarray}
&&  \frac{M_R -M_H}{m_Q} \sim Q \,,\qquad \qquad \frac{M_R - M_H }{M_H} \sim Q^2\,\qquad \qquad \;\;\, \;\,\,\,{\rm for}\qquad  H \parallel R \,,
\nonumber\\
&& \frac{M_{R}-M_{H}}{m_Q}  \sim Q^0 \,,\qquad \qquad \!\! \frac{|M_R - M_H| -  \Delta_H }{M_H}  \sim Q^2\,\,\,\qquad {\rm for} \qquad H \perp R\,,
\nonumber\\
&& \Delta_Q = \sqrt{ (M_H - M_R )^2- m_Q^2 }\,  \sim Q \qquad {\rm with}\qquad \Delta_H = \Delta\, M_H \lim_{m_{u,d,s}\to 0}\frac{1}{M_H}, \qquad
 \label{def-counting-D}
\end{eqnarray}
as properly formulated in terms of on-shell meson masses \cite{Lutz:2018cqo}. 
In (\ref{def-counting-D}) we use a notation $H \parallel R$ requesting  $H, R \in[0^-]$ or $H,R \in [1^-]$. Now, the bubble  $\Pi_H^{\rm bubble}$ can be decomposed in a convergent manner from the chiral limit up to the physical point.

\begin{figure}[t]
\center{
\includegraphics[keepaspectratio,width=0.49\textwidth]{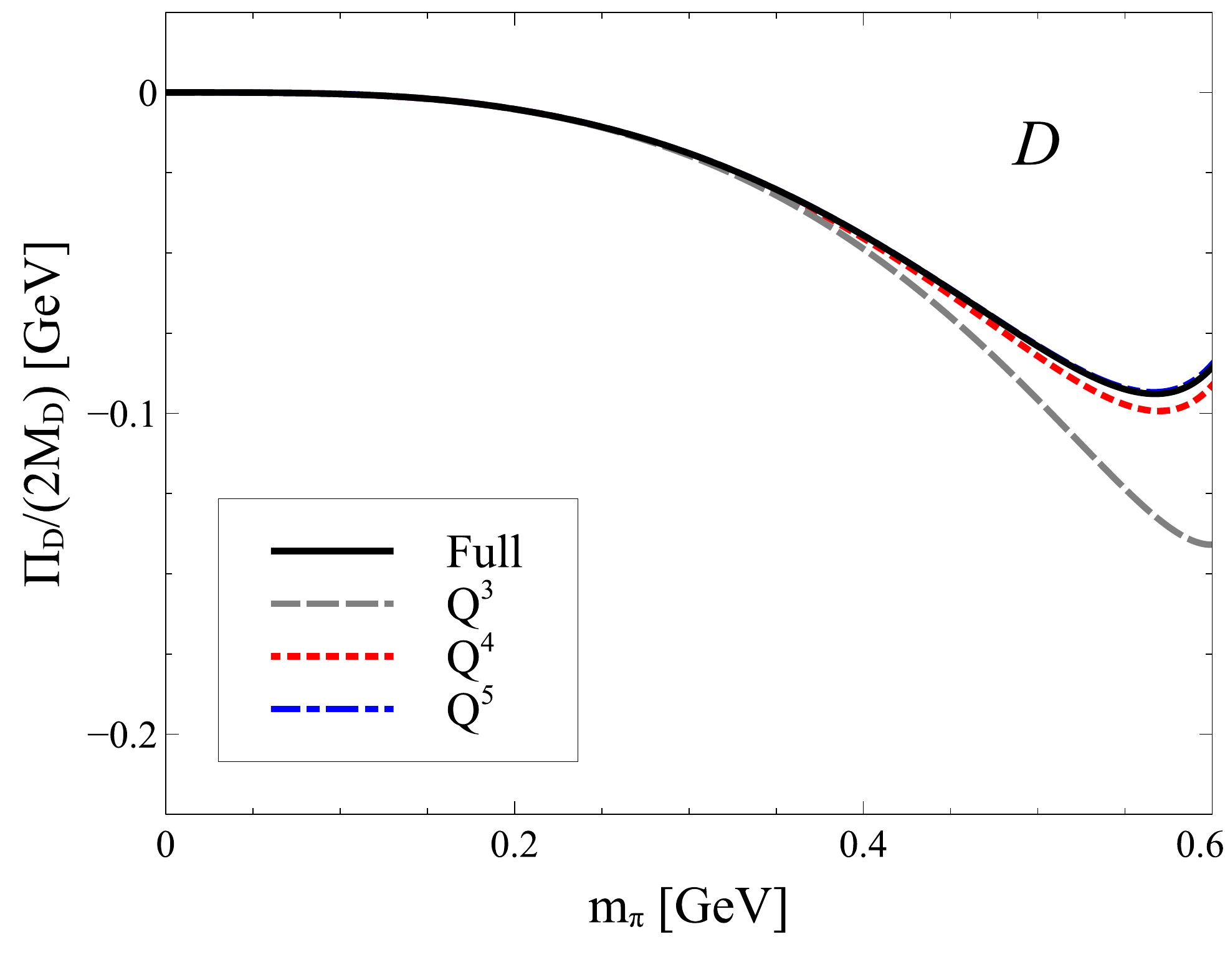}
\includegraphics[keepaspectratio,width=0.49\textwidth]{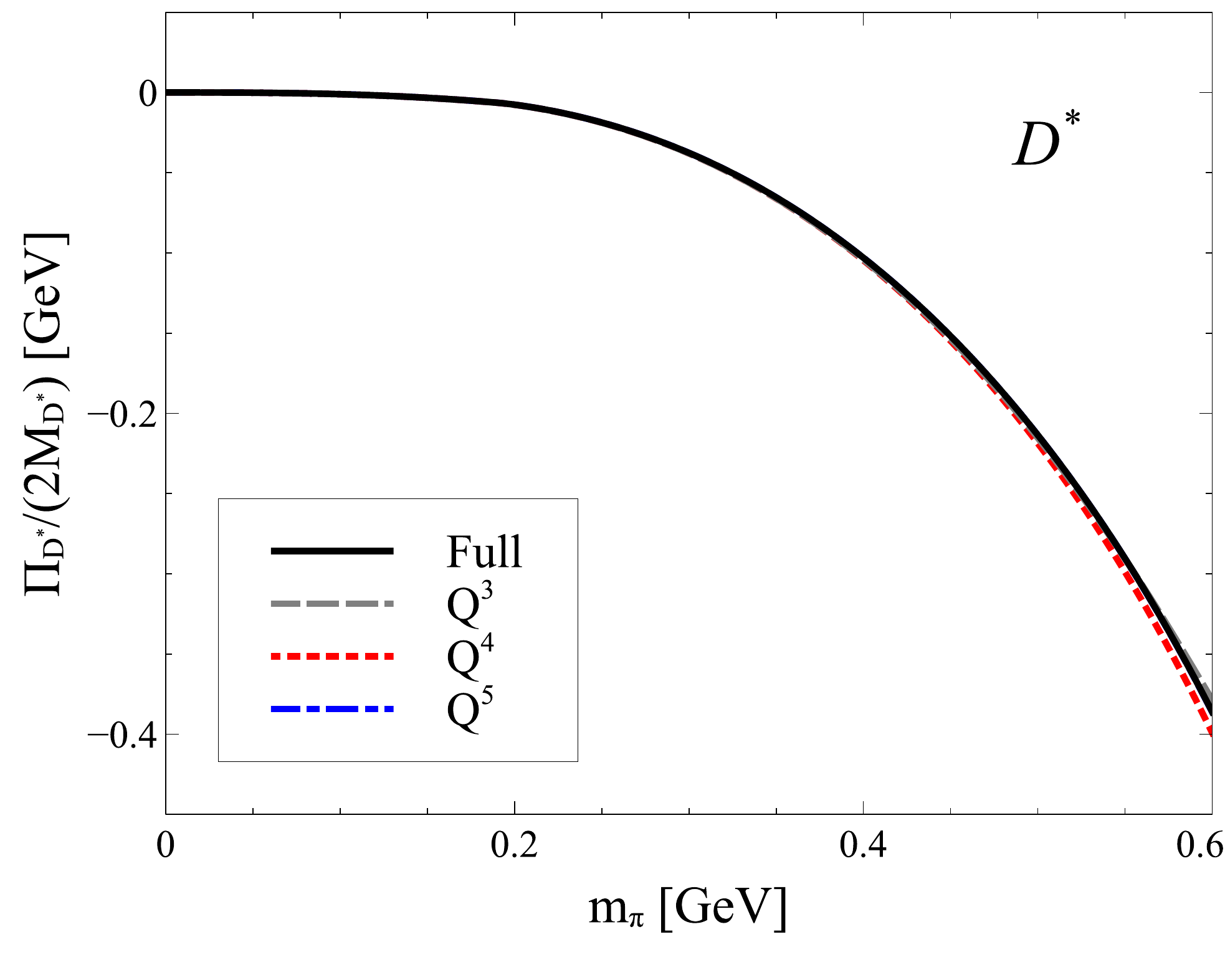}
}
\vskip-0.2cm
\caption{\label{fig:4} Charmed meson masses in the flavour limit as a function of the pion mass. The full result is shown by a solid line, while the truncated results up to $Q^3-Q^5$ according to the power counting \eqref{def-counting-D} are illustrated with broken lines.}
\end{figure}

\begin{table}[b]
\setlength{\tabcolsep}{3.5mm}
\renewcommand{\arraystretch}{1.2}
\begin{center}
\begin{tabular}{c|rrrrr }\hline
$H$             & $\bar \Pi^{\rm bubble }_H/(2\,M_H)$  & $\bar \Pi^{\rm bubble -3}_H/(2\,M_H)$ & $\bar \Pi^{\rm bubble -4}_H/(2\,M_H)$  & $\bar \Pi^{\rm bubble-5 }_H/(2\,M_H)$     \\ \hline \hline
$D$             &  -50.2  \, MeV                     &  -48.5  \, MeV                      &  -2.8  \, MeV                        &  1.1   \, MeV                           \\
$D_s$           &  -65.6 \, MeV                      &  -88.3 \, MeV                       &  20.1  \, MeV                        &  2.9   \, MeV                           \\
$D^*$           &  -113.4  \, MeV                    &  -99.5  \, MeV                      &  -17.1 \, MeV                        &  3.1   \, MeV                           \\
$D^*_s$         &  -166.1  \, MeV                    &  -197.5 \, MeV                      &  26.3  \, MeV                        &  6.6   \, MeV                           \\ \hline
\end{tabular}
\caption{The decomposition of the bubbles according to \eqref{def-counting-D}. We use $M = 1907.4$\,MeV and  $\Delta = 191.7$\,MeV.}
\label{tab:0}
\end{center}
\end{table} 

This is illustrated in the flavour limit with Fig. \ref{fig:4}. In comparison with the corresponding Fig. \ref{fig:2} a much improved convergence pattern is observed. 
Within the range  $0\leq m_\pi< 600$\,MeV, a quantitative reproduction of the bubble loop (solid line) is obtained. For the case where all the masses take their physical values, the 3rd, 4th and 5th components of the bubble loop  for $D$ and $D^*$ are listed in Tab.~\ref{tab:0}. We observe that the effects higher than $O(Q^4)$ are of a few MeV only.

\section{Fit to QCD lattice data}
   
We consider lattice results for charmed meson masses from 5 lattice groups.
On ETMC, PACS-CS and HSC ensembles pseudoscalar and vector $D$-meson masses are available \cite{Kalinowski:2015bwa,Mohler:2011ke,Lang:2014yfa,Moir:2016srx}. In contrast, only pseudoscalar masses are available on the lattice setups from HPQCD and LHPC \cite{Na:2012iu,Liu:2012zya}. On each lattice ensemble we solve the set of four coupled and non-linear equations 
where finite volume effects are taken into account \cite{Lutz:2014oxa}.
A non-standard scale setting is performed by the request that the four $D$-mesons reach their isospin averaged empirical values at the physical point.
In order to reduce the impact of a possible offset in the charm quark mass and 
discretization effects we consider only the mass splittings of the charmed mesons in our fit. The residual systematic error for the charmed-meson masses of about 5 MeV was estimated by the condition such that the $\chi^2$ per data point turns close to one. It is added in quadrature to the statistical error as given by the lattice groups.

On a given ensemble the quark masses, $m= (m_u+ m_d)/2$ and $m_s$, are determined from the lattice values of the pion and kaon masses. For Gasser and Leutwyler constants $L_4 -2\,L_6$ and $L_5 -2\,L_8$  we derived particular estimates in \cite{Guo:2018kno}. Our physical quark-mass ratio $m_s/m$ is compatible with the latest result of ETMC \cite{Carrasco:2014cwa} with $m_s/m = 26.66(32)$. Our ratios compare well with lattice results off the physical point in the few cases where they are available even though they did not enter our 
chisquare function \cite{Guo:2018kno}

\begin{table}[t]
\setlength{\tabcolsep}{2.5mm}
\renewcommand{\arraystretch}{1.1}
\begin{center}
\begin{tabular}{l|rrrr}
                                            &  Fit 1     &  Fit 2    & Fit 3    & Fit 4      \\ \hline
$ M\;\;$ \hfill [GeV]                       &  1.8762    &  1.9382   &  1.9089  & 1.8846  \\ \hline

$ c_0$                                      &  0.2270    &  0.3457   &  0.2957  &  0.3002  \\
$ c_1$                                      &  0.6703    &  0.9076   &  0.8765  &  0.8880  \\
$ c_2$                                      & -0.6031    & -2.2299   & -1.6630  & -1.3452  \\
$ c_3$                                      &  1.2062    &  4.5768   &  3.3260  &  3.0206  \\
$ c_4$                                      &  0.3644    &  2.0012   &  1.2436  &  0.9122  \\    
$ c_5$                                      & -0.7287    & -4.1445   & -2.4873  & -2.1393  \\    \hline
$ g_1\;\;$\hfill [GeV$^{-1}$]               &  0         &  0        &  0.4276  &  0.4407  \\
$ g_2\;\;$\hfill [GeV$^{-1}$]               &  0         &  0        &  1.0318  &  0.8788  \\
$ g_3\;\;$\hfill [GeV$^{-1}$]               &  0         &  0        &  0.2772  &  0.2003  

\end{tabular}
\caption{The low-energy constants (LEC) from four fit scenarios as explained in \cite{Guo:2018kno}. Each parameter set reproduces the isospin average of the empirical $D$-meson and $D^*$-meson masses from the PDG. 
The value $f = 92.4$ MeV was used in \cite{Guo:2018kno}.
}

\label{tab:1}
\end{center}
\end{table}

In Tab.\,\ref{tab:1} we recall four sets of LEC from  \cite{Guo:2018kno}. All four scenarios include not only the charmed meson masses, but also the s-wave scattering lengths \cite{Liu:2012zya}. In addition, Fit 2-4 are adjusted to the s-wave $\pi D$ phase shifts on a  HSC ensemble\cite{Moir:2016srx}. Fit 1 and 3 impose additional constraints from large-$N_c$ QCD with $c_2 = -c_3/2$ and $c_4 = -c_5/2$. 
Given the LEC of Tab. \ref{tab:1} we computed the coupled-channel s-wave scattering amplitudes in all isospin-strangeness sectors $(I,S)$. In this contribution we focus on the sector with $(I,S)= (0,1)$. 
A most remarkable prediction of chiral dynamics is the formation of the $D_{s0}^*(2317)$ as a coupled-channel $KD$ and $\eta D_s$ state \cite{Kolomeitsev:2003ac}. The $s$-channel unitarity is implemented according to \cite{Lutz:2001yb,Lutz:2003fm}. This approach relies on a renormalization condition where the unitarized amplitude matches the coupled-channel interaction kernel at a given matching scale $\mu_M$. If it is chosen to be close to the center of the
Mandelstam triangle s- and u-channel unitarized amplitudes can be matched smoothly in the vicinity of the matching point $\mu_M$ as is expected from the crossing symmetry condition. Small variations around the natural value of $\mu_M$ as suggested in \cite{Kolomeitsev:2003ac} may be used to access the uncertainty in the unitarization process. Given such a framework  the mass of $D_{s0}^*(2317)$ is well predicted with a rather small uncertainty in its mass even in a leading order computation. 

On the other hand, the isospin-violating width of $D_{s0}^*(2317)$ via the decay process $D_{s0}^*(2317) \to \pi^0 D_s$  depends sensitively on details of the dynamical scenario \cite{Faessler:2007gv,Lutz:2007sk}.
The leading order Weinberg-Tomozawa interaction suggests the width to be 75\,keV \cite{Lutz:2007sk}. A first estimate of the impact of chiral corrections suggested a much larger width of 140\,keV \cite{Lutz:2007sk}. An improved estimate of $(133\pm 22)$\,keV is based on first lattice results on some s-wave lengths lengths \cite{Liu:2012zya}. Given our fit scenarios we confirm that the mass of $D_{s0}^*(2317)$ is recovered within a small variation of the natural matching scale $\Delta \mu_M = \pm 0.1$\,GeV for all the four fits. 
We emphasize that in none of the four chisquare functions  we used in our fits the mass of the $D_{s0}^*(2317)$ entered. 
In Tab. \ref{tab:2}, we display  the predictions of the hadronic width of the $D_{s0}^*(2317)$ from the four fits  \cite{Guo:2018kno}.
The results depend on the choice of the $\pi^0-\eta$ mixing angle $\epsilon$. 
While in the previous study \cite{Lutz:2007sk}, the value $\epsilon = 0.010(1)$ was used, the recent lattice study suggests $\epsilon = 0.0122(18)$\cite{Carrasco:2014cwa}. We provide predictions for both values of $\epsilon$ in Tab. \ref{tab:2}.
With $\epsilon = 0.0122(18)$, we arrive at our estimate for the width of $D_{s0}^*(2317)$ as $(104-116)$\,keV, taking into account the results for the $\pi D$ scattering phase shifts on a HSC ensemble.

\begin{table}[t]
\setlength{\tabcolsep}{2.5mm}
\renewcommand{\arraystretch}{1.2}
\begin{center}
\begin{tabular}{l|cccc| c} 
                                                   &  Fit 1    &  Fit 2    &  Fit 3   &  Fit 4  & $\epsilon $ \\ \hline
                                                      
$\Gamma_{D^*_{s0}(2317)\to \pi^0 D_s }$    [keV]   &   61.1    & 54.1      & 88.6     & 80.1    & 0.0100  \\
                                                   &  74.6     &  68.4     &  115.8   & 104.4   & 0.0122

\end{tabular}
\caption{Prediction for the isospin violating decay width of the $D^*_{s0}(2317)$ in the four fit scenarios of Tab. \ref{tab:1}. }
\label{tab:2}
\end{center}
\end{table}

\section{Summary}

We studied the chiral extrapolation of charmed meson masses based on the three-flavour chiral Lagrangian. It was illustrated that once the chiral expansion is organized in terms of on-shell meson masses a well convergent expansion is obtained that can be applied from the chiral limit to the physical point faithfully. The framework was applied to the data on charmed meson masses with $J = 0^-$ and $J^P = 1 ^-$ based on 
5 different lattice setups from  ETMC, PACS-CS, HSC,  HPQCD and LHPC.  
Additional constraints from some s-wave scattering lengths within the LHPC setup and first results on a HSC ensemble for the $\pi D$ and $  \eta\, D$ scattering phase shifts were imposed. Four sets of low-energy constants were discussed, two of them provide an excellent reproduction of all considered lattice data.

The implication of higher order counter terms in the coupled-channel dynamics involving $D$-mesons was discussed at hand of the $(I,S)=(0,1)$ sector. 
Our predicted range for the isospin violating decay width of the $D_{s0}^*(2317)$ is  $(104-116)$\,keV. This magnitude is within the expected resolution of the PANDA experiment at FAIR. A measurement of this width is important since it provides more insight into the chiral dyanmics of the open-charm sector of QCD. The size of the chiral correction terms is crucially linked  to the fate of possible flavour exotic open-charm meson states.

\bibliographystyle{JHEP}
\bibliography{thesis}
\end{document}